\documentclass[aps,prl,reprint,twocolumn,superscriptaddress,amsmath,amssymb,citeautoscript,notitlepage]{revtex4-1}
\usepackage{amsmath}
\usepackage{amssymb}
\usepackage{graphicx}
\usepackage{color}
\usepackage{bm}
\usepackage{float}
\usepackage{enumitem}

\usepackage[normalem]{ulem}
\usepackage[section]{placeins}

\usepackage{epstopdf}
\usepackage{grffile}
\DeclareGraphicsExtensions{.eps}

\newcommand{\e}{{\rm e}}
\newcommand{\ii}{{\rm i}}

\newcommand{\la}{{\langle}}
\newcommand{\ra}{{\rangle}}

\usepackage{xspace}

\begin{document}

\title{Universal Hall Response in Synthetic Dimensions}
\author{Sebastian Greschner}
\author{Michele Filippone}
\author{Thierry Giamarchi}
\affiliation{Department of Quantum Matter Physics, University of Geneva, 1211 Geneva, Switzerland}

\begin{abstract}
We theoretically study the Hall effect on interacting $M$-leg ladder systems, comparing different measures and properties of the zero temperature Hall response in the limit of weak magnetic fields. Focusing on $SU(M)$ symmetric interacting bosons and fermions, as relevant for {\it e.g.} typical synthetic dimensional quantum gas experiments, we identify an extensive regime in which the Hall imbalance $\Delta_{\rm H}$ is universal and corresponds to a classical Hall resistivity $R_{\rm H}=-1/n$ for a large class of quantum phases. Away from this high symmetry point we observe interaction driven phenomena such as sign reversal and divergence of the Hall response.
\end{abstract}
\date{\today}
\maketitle

%%%%%%%%%%%%%%%%%%%%%%%%%%%%%%%%%%%%%%%%%%%%%%%%%%%%%%%%%%%%%%%%%%%%%%%%%%%%%%%%%%%%%%%%%%%%%%%%%%%%%%
%%%% INTRO
%%%%%%%%%%%%%%%%%%%%%%%%%%%%%%%%%%%%%%%%%%%%%%%%%%%%%%%%%%%%%%%%%%%%%%%%%%%%%%%%%%%%%%%%%%%%%%%%%%%%%%

In its semi-classical approximation~\cite{AshcroftMermin} the Hall response of a conductor threaded by a weak perpendicular magnetic field $B$ is independent of the bar geometry. Due to Galilean invariance, the ratio of the electric field $E_y$ and the longitudinal current density $j_x$, the Hall coefficient $R_{\rm H} = E_y / j_x B$, uniquely depends on the effective charge-$q$  carrier density $n$, $R_{\rm H}=-1/n q$, providing an extraordinary tool for  the characterization of  solid state systems~\cite{popovic2003hall,chien2013hall, Badoux2016}. Nevertheless, deviations from parabolic bands in realistic condensed matter systems lead dependence on the curvature of the Fermi surface~\cite{Tsuji1958, Ong1991, Haldane2005} and large deviations of the Hall coefficient from its classical expressions are expected in strongly correlated phases, in particular when constrained to low dimensions, {\it e.g.} ladder-like systems. 
Several theoretical approaches addressed  the Hall effect in strongly correlated quantum phases \cite{Brinkman1971,Shastry1993, lange1997memory,Prelovsek1999,*Zotos2000,Auerbach2018,Lopatin2001, Leon2007,Huber2011, Berg2015}, but unbiased calculations of the Hall coefficient remain challenging in interacting systems. 

Tremendous experimental progress with ultracold lattice gases~\cite{Aidelsburger2013,Miyake2013,Atala2014,Tai2016,Goldman2016} in artificial magnetic fields paves the way  to the exciting study of the Hall effect in highly controllable clean many-body systems. 
Several experiments have so far observed the Hall~\cite{Aidelsburger2013,Stuhl2015, Mancini2015} and quantum Hall~\cite{Atala2013,Aidelsburger2015} effect and a systematic measurement of the Hall coefficient was shown recently by Genkina~{\it et al.}~\cite{Genkina2018}. Several of these experiments were performed with synthetic-lattice dimensions~\cite{Celi2014, Stuhl2015, Mancini2015,Livi2016,Kolkowitz2016,Alex2017}, realizing Harper-Hofstadter (HH) like models on a ladder~\cite{harper55,Hof1976}. Promising ongoing efforts towards the realization of strong correlations in such systems~\cite{Tai2016, Alex2018} motivate the detailed theoretical analysis of the Hall effect in interacting many body systems.

\begin{figure}[t]
\centering
\includegraphics[width=1\linewidth]{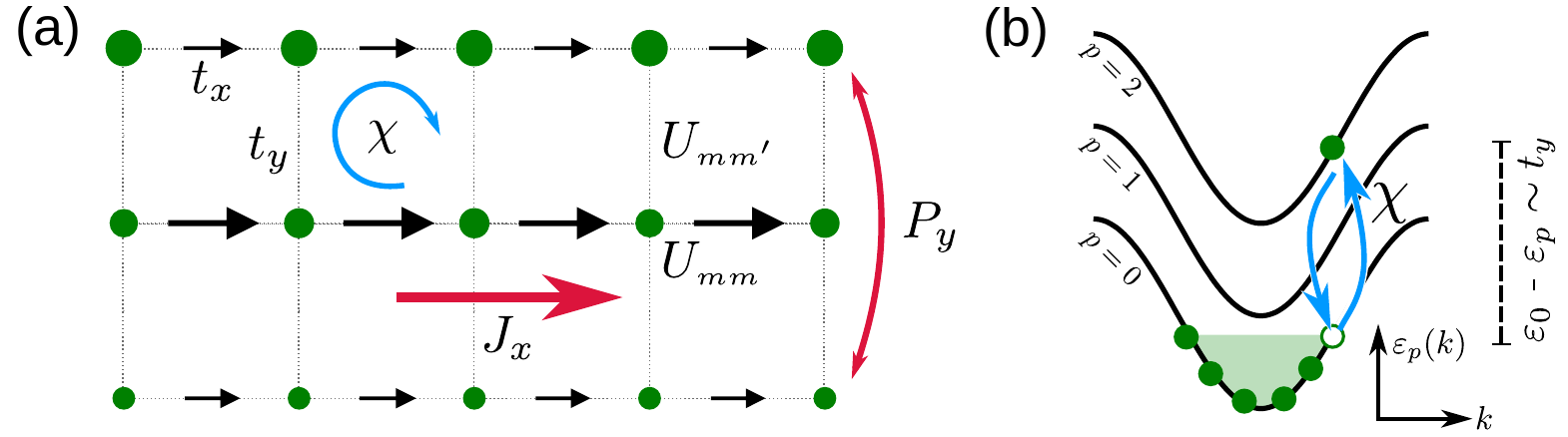}
\includegraphics[width=0.49\linewidth]{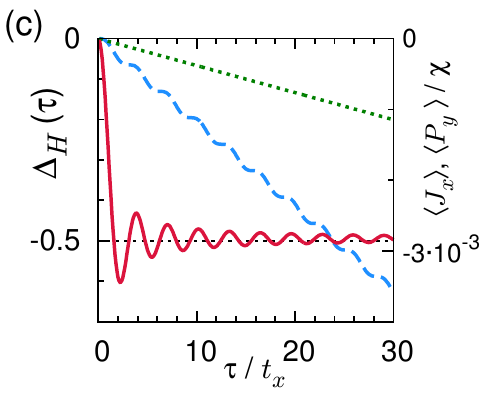}
\includegraphics[width=0.49\linewidth]{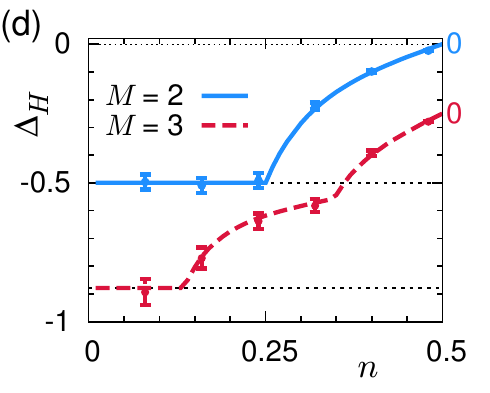} 
\caption{ {
(a) Particle (dots) and  current (arrows) density on an $M$-leg ladder threaded by a flux $\chi$, after applying a tilt ($M=3$, subtracting $\tau=0$ values for clarity). 
(b) Band structure of the $M$-leg ladder as function of the wave-vector $k$ along the $x$-direction. 
(c) Quench dynamics of the Hall imbalance $\Delta_{\rm H}$, polarization $\la P_y\ra / \chi$ and current $\la J_x\ra$ (solid, dashed and dotted lines respectively) for tilted free fermions ($t_x=t_y$, $n=1/5$, $\chi=0.01$, $\Delta\mu=0.01$, see text).
(d) $\Delta_{\rm H}$ for free fermions as function of  $n$. Solid lines correspond to Eq.~\eqref{eq:DeltaHfreefermions}, and points with error bars to time averages of the dynamics in (c). $\Delta_{\rm H}$ is constant when only the lower band is occupied and kinks correspond to the occupation of upper bands in (b).}}
\label{fig:1}
\end{figure}

In this Letter, we study the Hall response of strongly interacting fermions and bosons on quantum ladders. We introduce the Hall imbalance $\Delta_{\rm H}$, which can be directly observed in ultracold atom experiments. The key result is the observation of an extensive {\it universal} regime of parameters, where $\Delta_{\rm H}$ is constant, corresponding to $R_{\rm H}=-1/n$ behavior, independent of particle statistics and interaction strength.
By means of numerical matrix product state DMRG simulations~\cite{White1992,Schollwoeck2011}, supported by analytical arguments, we analyze the robustness of this effect and show  its relevance for quench dynamics experiments with state-of-the-art quantum gases with synthetic dimensions. We discuss the breaking of this universality out of $SU(M)$ symmetry, in which  divergencies or sign reversals of $\Delta_{\rm H}$ signal phase transitions. We also provide an approximate formula to calculate $\Delta_{\rm H}$ at equilibrium with open boundary conditions.

We consider HH ribbons  on $M$-legs, see Fig.~\ref{fig:1}(a). The Hamiltonian is
$H=H_{\rm kin}^x + H_{\rm kin}^y +H_{\rm int}$,
with hopping along the ladder
$H_{\rm kin}^x = -t_x \sum_{j,m} \e^{\ii\chi \cdot (m-m_0)} a^\dagger_{j,m} a_{j+1,m} + {\rm H.c.}$ ($m_0=(M-1)/2$ and $m\in[0,M-1]$)
and in the transverse direction $H_{\rm kin}^y = - t_y \sum_{j,m} a^\dagger_{j,m} a_{j,m+1} + {\rm H.c.}$.
Here, $a_{j,m}^{(\dagger)}$ is a fermionic or bosonic annihilation(creation) operator on the ladder rung $j$ and leg $m$. In this work, we consider on-rung interactions as relevant for synthetic dimension experiments
$H_{\rm int} = \sum_{j,m,m'}\frac{U_{mm'}}{2} n_{j,m} n_{j,m'}$, with $n_{j,m}=a_{j,m}^\dagger a_{j,m}$.
Typical ultracold atoms experiments  realize an approximate $SU(M)$ symmetry $U_{m,m'} = U$~\cite{SUMexperimentnote}\nocite{Bloch2008,Cazalilla2014}.

%%%%%%%%%%%%%%%%%%%%%%%%%%%%%%%%%%%%%%%%%%%%%%%%%%%%%%%%%%%%%%%%%%%%%%%%%%%%%%%%%%%%%%%%%%%%%%%%%%%%%%
%%%%%  REACTIVE ALL RESPONSE 
%%%%%%%%%%%%%%%%%%%%%%%%%%%%%%%%%%%%%%%%%%%%%%%%%%%%%%%%%%%%%%%%%%%%%%%%%%%%%%%%%%%%%%%%%%%%%%%%%%%%%%

Recent experiments, such as Ref.~\cite{Genkina2018}, accelerate the lattice gas by a linear potential $\Delta \mu \sum_{jm} j n_{jm}$.
Subsequent monitoring of the evolution of the spin resolved momentum distribution allows for the measurement of the density polarization $P_y = \sum_{m=0}^M (m-m_0) n_{j, m}$  and the current $J_x = -\ii t_x \sum_{j,m} \e^{\ii\chi \cdot (m-m_0)} a^\dagger_{j,m} a_{j+1,m} + {\rm H.c.}$ as function of time $\tau$.  
Figures~\ref{fig:1}(a-c) sketch this procedure for a $M=3$ leg ladder system, initially  prepared in the ground state.  
After the quench for $\tau>0$ a total current develops $\la J_x \ra\neq 0$  as well as a finite density imbalance $\la P_y \ra \neq 0$. Both quantities essentially grow linearly with $\tau$ and, hence, the resulting Hall imbalance 
\begin{align}\label{eq:dh}
\Delta_{\rm H} = \left.\frac{\la P_y \ra}{\chi \cdot \la J_x \ra}\right|_{\chi\rightarrow 0}
\end{align}
oscillates around a finite constant value for small enough times $\tau \lesssim L/2t$ as long as the finite size of the system can be neglected, see Fig.~\ref{fig:1}(c).

%%%%%%%%%%%%%%%%%%%%%%%%%%%%%%%%%%%%%%%%%%%%%%%%%%%%%%%%%%%%%%%%%%%%%%%%%%%%%%%%%%%%%%%%%%%%%%%%%%%%%%
In case of adiabatic dynamics, the Hall imbalance~\eqref{eq:dh} can be equivalently derived {\it at equilibrium} for the same system on a ring with $L$ rungs threaded by a Aharonov-Bohm flux $\phi$, {\it i.e.} with periodic boundary conditions (PBC) along $x$ and the substitution $a^\dagger_{j,m}a_{j,m+1} \rightarrow e^{i\phi/L}a^\dagger_{j,m}a_{j,m+1}$~\cite{Adiabaticnote}.  The flux $\phi$  induces a persistent current $\la J_x \ra = L\,\partial_\phi \la H \ra \big|_{\phi\to 0}$,  leading to a finite imbalance \eqref{eq:dh} for $\chi\neq0$. 
This {\it reactive Hall response} has been discussed in detail by Prelov\v sek {\it et al.}~\cite{Prelovsek1999,*Zotos2000}: the Hall coefficient  $R_{\rm H}$ is found by adding an extra term $E_y P_y$ to the Hamiltonian, adjusted such that $\la P_y\ra = 0$~\cite{Prelovsek1999,supmat}\nocite{filippone2018controlled,kohn64,shastry90,millis90,gu2010}. Nevertheless, the Hall imbalance $\Delta_{\rm H}$ is a more direct and simpler observable in recent quantum gas experiments and we will consider both quantities in the following.  Remarkably, both  $\Delta_{\rm H}$ and $R_{\rm H}$ can be expressed as derivatives of the ground state energy $\mathcal E_0(\phi,\chi,E_y)$~\cite{supmat},
$L \Delta_{\rm H}= \partial_{\phi\chi E_y} \mathcal E_0 / \partial_{\phi\phi} \mathcal E_0$ and $R_{\rm H}=- L\Delta_{\rm H} / \partial_{E_y E_y} \mathcal E_0$ for $\phi,\chi,E_y=0$.

%%%%%%%%%%%%%%%%%%%%%%%%%%%%%%%%%%%%%%%%%%%%%%%%%%%%%%%%%%%%%%%%%%%%%%%%%%%%%%%%%%%%%%%%%%%%%%%%%%%%%%
%%%%%%%% FREE PARTICLES 
%%%%%%%%%%%%%%%%%%%%%%%%%%%%%%%%%%%%%%%%%%%%%%%%%%%%%%%%%%%%%%%%%%%%%%%%%%%%%%%%%%%%%%%%%%%%%%%%%%%%%%
We first identify the universal regime of interest shown in Fig.~\ref{fig:1}, in which $\Delta_{\rm H}$ is a constant function of $n$, corresponding to $R_{\rm H}=-1/n$, which can be well understood in free particle systems. For $\chi,E_y=0$, the generic spectrum of $M$ coupled wires on the lattice coupled by hopping to each other is made of $M$ bands $\varepsilon_p(k)=\varepsilon_x(k)+\varepsilon_y(p)$, that are labeled with the index $p\in[0,M-1]$, in $k$ space (wave-vector along $x$-direction). As shown in Fig.~\ref{fig:1}(b) the bands are split by the transverse hopping strength $t_y$. In the specific case of the HH model we have $\varepsilon_{x}(k)=-2t_{x}\cos(k)$ and $\varepsilon_{y}(p)=-2t_{y}\cos(\pi (p+1)/(M+1))$. Analytical expressions of $\Delta_{\rm H}$ are readily found in perturbation theory in $\chi,\,E_y,\phi$~\cite{supmat}. For non-interacting fermions one finds
\begin{align}
\Delta_{\rm H}&=\frac{\sum_{p<P} v_{F,p} \mathcal{I}_{p} }{ \sum_{p<P} v_{F,p} }\,, & R_{\rm H}=\frac{-\Delta_{\rm H} }{ \sum_{p<P} n_p\mathcal I_p }\,.
\label{eq:DeltaHfreefermions}
\end{align}
in which $v_{F,p}$ and $n_p$ are the Fermi velocity and density of fermions in band $p$ and $P\leq M$ is the number of occupied bands, see Fig.~\ref{fig:1}(b). The coefficients $\mathcal{I}_{p}$ depend on the details of the Hamiltonian~\cite{supmat}. 
Figure~\ref{fig:1}(d) shows examples of $\Delta_{\rm H}$ as a function of the total density $n=\sum_pn_p$, from Eq.~\eqref{eq:DeltaHfreefermions}. The Hall imbalance  $\Delta_{\rm H}$ exhibits a series of kinks, corresponding to the change in the number of occupied bands $P$. It is remarkable, as shown in panels~(c-d) of Fig.~\ref{fig:1}, that non adiabatic behavior (even though $\Delta\mu$ is a ``weak'' perturbation) is perfectly reproduced by equilibrium calculations. Due to particle-hole symmetry the Hall response vanishes identically at half filling~\cite{Lopatin2001} and for the case of hole-conductance $n>0.5$ the sign of the Hall imbalance is inverted~\cite{Prelovsek1999,*Zotos2000}. The Hall response~\eqref{eq:DeltaHfreefermions} is finite when only the lowest band $p=0$ is occupied:
\begin{align}
\Delta_{\rm H} = q\mathcal{I}_0(M, t_\perp) \text{ and } R_{\rm H} = -1/n \,.
\label{eq:DeltaHSUM}
\end{align}
This is an interesting result as it recovers {\it exactly} the usual $1/n$ behavior for $R_{\rm H}$ of free particles in the continuum, generally violated on the lattice~\cite{Tsuji1958, Ong1991, Haldane2005} and {\it always} applies   for non-interacting bosons as well~\cite{supmat}. Note, that Eq.~\eqref{eq:DeltaHSUM} shows possibility to observe a finite Hall response in systems in which only two Fermi points are present, for which one would naively have expected a single chain behavior and thus the absence of Hall effect. 
This paper focuses on the regime corresponding to Eq.~\eqref{eq:DeltaHSUM}. We will now provide numerical calculations supported by analytical arguments that this observation carries over to the correlated regime for generic $SU(M)$ symmetric bosons or fermions in a large family of ground states.

%%%%%%%%%%%%%%%%%%%%%%%%%%%%%%%%%%%%%%%%%%%%%%%%
%%%%%%%%%%%%%%%%%%%%%%%%%%%%%%%%%%%%%%%%%%%%%%%%
%%%%%%% NUMERICS INTERACTIONS
%%%%%%%%%%%%%%%%%%%%%%%%%%%%%%%%%%%%%%%%%%%%%%%%
%%%%%%%%%%%%%%%%%%%%%%%%%%%%%%%%%%%%%%%%%%%%%%%%

\begin{figure*}[t]
\centering
\includegraphics[height=4.7cm]{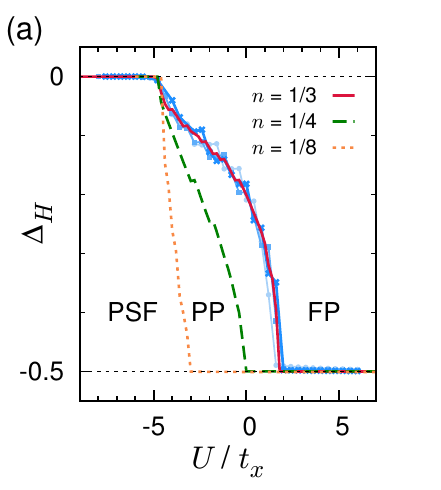}
\includegraphics[height=4.7cm]{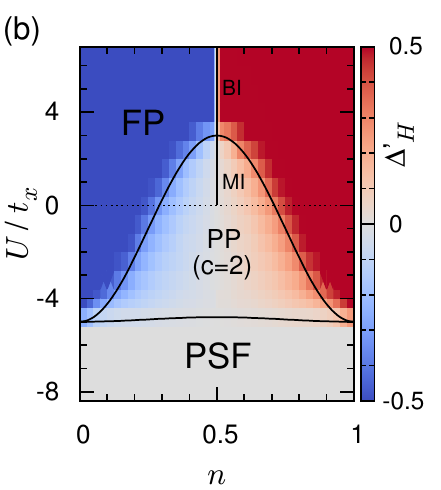}
\includegraphics[height=4.7cm]{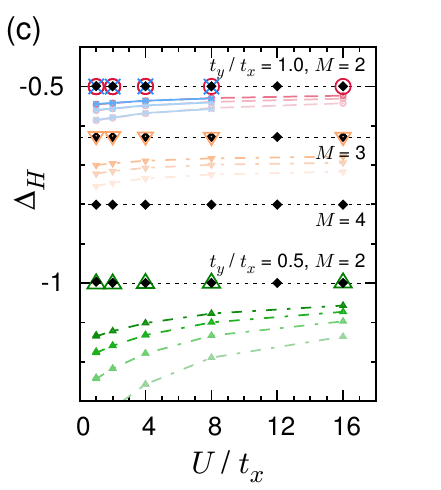}
\includegraphics[height=4.7cm]{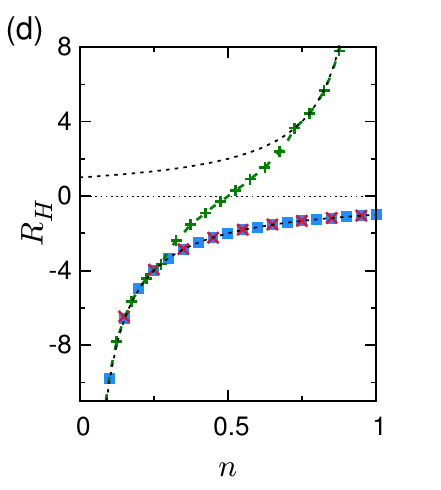}
\caption{ {Hall effect in $SU(M)$ symmetric ladders. 
(a) $M=2$ fermions with on-rung interactions $U$ (DMRG data, $t_x=t_y$) for different fillings $n=1/3$, $1/4$ and $1/8$. Symbols depict PBC results for $L=24,36,48$ ($\circ,\Box,\times$) rungs for $n=1/3$. The other lines correspond to $\Delta_{\rm H}'$ (OBC, $L=96$ rungs).
(b) Color-plot of $\Delta_{\rm H}$ for interacting fermions as in (a). The solid lines correspond to the phase transitions for the one dimensional Fermi-Hubbard model at $\chi=0$ with Zeeman field of strength $t_y$~\cite{Essler2005}.
(c) $\Delta_{\rm H}$ of the $SU(M)$-symmetric Bose Hubbard model for $M=2$ and $t_x=t_y$, $n=1/2$ and $n=1/4$ as well as $t_x=t_y/2$, $n=1/4$  (small symbols: $\times,\circ,\triangle$) and $M=3$, $t_x=t_y$, $n=1/8$ ($\triangledown$) with (bottom to top) $L=24,28,32$, and $36$ rungs - larger symbols depict the extrapolated value in the thermodynamic limit $L\to\infty$~\cite{supmat}. 
The filled $\diamond$ symbols show $\Delta_{\rm H}'$ data for $M=2$ and $3$ and $4$ leg ladders ($L=60$, $n=1/2M$). The horizontal dashed lines depict Eq.~\eqref{eq:DeltaHSUM}.
(d) $1/n$-behavior (dotted lines) of the Hall resistivity $R_{\rm H}$ for $M=2$ bosons with $U=16t_x$, $t_x=t_y$ ($\times$) and $U=8t$, $t_x=t_y/2$ ($\Box$) as well as fermions $U=t_x=t_y$ ($+$). 
}
}
\label{fig:2}
\end{figure*}

Figures~\ref{fig:2}(a-b) show  DMRG results for $\Delta_{\rm H}$ for interacting fermions on $M=2$ leg ladder as function of $n$ and $U$ (see the Supplemental Material~(SM)~\cite{supmat} for further examples). 
The solid lines of Fig.~\ref{fig:2}(b) indicate the different phase transitions for $\chi=0$ known for the integrable Fermi-Hubbard model for finite $t_y$. In the fully polarized (FP) state, for which all $N$ particles occupy the lowest band $p=0$, we find $\Delta_{\rm H} = - 1/2$ is independent of interaction strength and density in perfect accordance with Eq.~\eqref{eq:DeltaHSUM}.

Due to particle-hole symmetry in the Mott (MI) and band insulator (BI) phases at half filling $\Delta_{\rm H}$ vanishes identically. As the fully paired phase (PSF),  superfluid of composite bosonic pairs, for attractive interactions can be related to the insulating MI phase by means of a Shiba transformation~\cite{Essler2005}, we also find here identically vanishing $\Delta_{\rm H}$. 
In the partially paired~(PP) two component phase (with a central charge $c=2$), as in the  non-interacting case, $\Delta_{\rm H}$ strongly deviates from the universal form interpolating smoothly between results of the FP and PSF phases. 

Remarkably, this universal behavior carries over to interacting bosons. Figure~\ref{fig:2}(c) shows the universal values for softcore $SU(M)$ symmetric bosons for $M=2, 3$ and $4$, different densities and interactions strengths. As anticipated, $\Delta_{\rm H}$ is just given by Eq.~\eqref{eq:DeltaHSUM}, that is a function of $t_y/t_x$, independent of the interaction strength. 
Note, that $J_x$ and $P_y$ themselves  exhibit a complicated dependence on the parameters $n$ and $U$.
Interestingly, for the bosonic model, as it is not particle-hole symmetric, we observe approaching of the same finite constant even for half filling, where for sufficiently strong interactions the system enters an insulating state. In order to verify the universality of the corresponding Hall coefficient, in Fig.~\ref{fig:2}(d) we show $R_{\rm H}$ for some of the previous examples. As conjectured we find $R_{\rm H} = -1/n$ for bosons and fermions independent of $M$, density and interaction strength as long as only the lower band $p=0$ is occupied.

%%%%%%%%%%%%%%%%%%%%%%%%%%%%%%%%%%%%%%%%%%%%%%%%%%%%%%%%%%
%%%%%%%%%%%%%%%%%%%%%%%%%%%%%%%%%%%%%%%%%%%%%%%%%%%%%%%%%%
%%%%%%%%%%% ANALYTICAL EXPLANATION
%%%%%%%%%%%%%%%%%%%%%%%%%%%%%%%%%%%%%%%%%%%%%%%%%%%%%%%%%%
%%%%%%%%%%%%%%%%%%%%%%%%%%%%%%%%%%%%%%%%%%%%%%%%%%%%%%%%%%

The universality of Eq.~\eqref{eq:DeltaHSUM} is understood by inspecting  the expectation values of $P_y$ and $J_x$ in general many-body perturbation theory. 
Upon introduction of  the eigenstates $\{|\alpha\rangle\}$  of $H(\chi=0)$, of energy $\mathcal E_\alpha$, with $|0\rangle$ the ground state, 
the leading contribution to the polarization  reads~\cite{supmat}
\begin{equation}\label{eq:pertDeltan}
\begin{split}
\la 0| P_y|0 \ra =& \chi \sum_{\alpha\neq 0}\frac{\la 0| P_y |\alpha \ra \la \alpha| \tilde{J}_x |0 \ra+\mbox{c.c.}}{\mathcal E_0 -\mathcal  E_\alpha}
\end{split}\,,
\end{equation}
in which we introduced the asymmetric current $\tilde{J}_x =-t_x\sum_{j, m} \ii \e^{\ii \phi/L} (m-m_0) a^\dagger_{j,m} a_{j+1,m}  + {\rm H.c.}$. Consider the commutator $[P_y, H(\chi=0)] = \sum_{j,p,p'} C_{p,p'} (\epsilon_p - \epsilon_{p'}) \tilde{a}_{j,p}^\dagger \tilde{a}_{j,p'}$ in which we switched to operators $a_{j,p}$ diagonalizing $H_{\rm kin}^y$ and thus annihilating particles on band $p$ (the factor $C_{p,p'}$ is given in the SM~\cite{supmat}).
We consider the FP ground state, stabilized {\it e.g.} by repulsion, large $t_y$, or bosonic enhancement. For excited states $|\alpha_p\rangle$, with $1$ particle in band $p>0$, the commutator leads to the fact that $[\mathcal E_{\alpha_p} -\mathcal  E_0] \la \alpha_p | P_y | 0 \ra = [\epsilon_y(p) - \epsilon_y(0)] \la \alpha_k | P_y | 0 \ra$. For the cases in which $\la \alpha_p | P_y | 0 \ra\neq 0$, the energy difference of the interacting many-body states becomes trivial $\mathcal E_{\alpha_p} - \mathcal E_0 = \epsilon_y(p) - \epsilon_y(0)$.
 As an important consequence,  the ground state polarization reads 
$\la P_y \ra = \chi \mathcal{I}_0 \la 0 | J_x |0\ra$~\cite{supmat},
leading to the remarkably simple expression Eq.~\eqref{eq:DeltaHSUM} for the Hall imbalance for any single  component $SU(M)$ symmetric quantum state on a $M$-leg ladder. The behavior $R_{\rm H}=-1/n$ follows by a similar argument to calculate $\partial^2\mathcal E_0/\partial E_y^2$ in perturbation theory~\cite{supmat}. Note, that for generic ladder models these results are generally true in the large coupling limit $t_y / t_x \gg 1$.

%%%%%%%%%%%%%%%%%%%%%%%%%%%%%%%%%%%%%%%%%%%%%%%%%%%%%%%%%%%%%%%%%
%%%%%%%%%%%%%%%%%%%%%%%%%%%%%%%%%%%%%%%%%%%%%%%%%%%%%%%%%%%%%%%%%
%%%%%%%%%%    WEAKLY INTERACTING BOSONS
%%%%%%%%%%%%%%%%%%%%%%%%%%%%%%%%%%%%%%%%%%%%%%%%%%%%%%%%%%%%%%%%%
%%%%%%%%%%%%%%%%%%%%%%%%%%%%%%%%%%%%%%%%%%%%%%%%%%%%%%%%%%%%%%%%%

We consider now breaking the $SU(M)$ symmetry. Note, that the Hall response~\eqref{eq:DeltaHSUM} for fermions in the lowest band is robust, since after diagonalization of $H_{\rm kin}^y$, $H_{\rm int}$ takes the general form $H_{\rm int}=\sum_{p,p',q,q'}\mathcal U_{p,p',q,q'}a^\dagger_{j,p}a_{j,p'}a^\dagger_{j,q}a_{j,q'}$, which projects to zero if only the lowest band $p=0$ is occupied.  This is not the case for bosons. 
Important insight into the deviations  of the bosonic Hall effect from Eq.~\eqref{eq:DeltaHSUM} may be obtained by a simple mean-field description, justified for typical experiments with large particle numbers per site~\cite{Genkina2018}. 
For small fields $\chi$ and $\phi$,   each site can be described by a coherent state with fixed density $n_{j,m}=n+(m-m_0)\delta n$, leading to a  classical description of the system~\cite{Struck2012,Struck2013}.  The density variation $\delta n$ is found by minimization of the energy $\la H \ra$ and the total current is given by the twist between subsequent sites $\la J_x \ra= 4 t_x n \sin(\phi/L) \approx 4 t_x n \phi/L$. The Hall imbalance $\Delta_{\rm H}$ is then derived and for $M=2$~\cite{supmat} 
\begin{align}
\Delta_{\rm H} = - \frac{t_x}{2 t_y + 2 n (U_{00} - U_{01})}\,.
\label{eq:DHcoh}
\end{align}
A rich phenomenology of the Hall effect is thus suggested. At the $SU(2)$ symmetric point, the interaction part of the ground-state energy simplifies to $U n^4$ and, hence, becomes independent of rung density variations $\delta n$. Thus $\Delta_{\rm H}$ remains independent of $U$ and $n$ obeying Eq.~\eqref{eq:DeltaHSUM}.

%%%%%%%%%%%%%%%%%%%%%%%%%%%%%%%%%%%%%%%%%%%%%%%%%%%%
%%%%%%%%%%%%%%%%%%%%%%%%%%%%%%%%%%%%%%%%%%%%%%%%%%%%
%%%%%%%%% ASYMMETRIC CASE 
%%%%%%%%%%%%%%%%%%%%%%%%%%%%%%%%%%%%%%%%%%%%%%%%%%%%
%%%%%%%%%%%%%%%%%%%%%%%%%%%%%%%%%%%%%%%%%%%%%%%%%%%%

\begin{figure}[tb]
\centering
\includegraphics[width=1\linewidth]{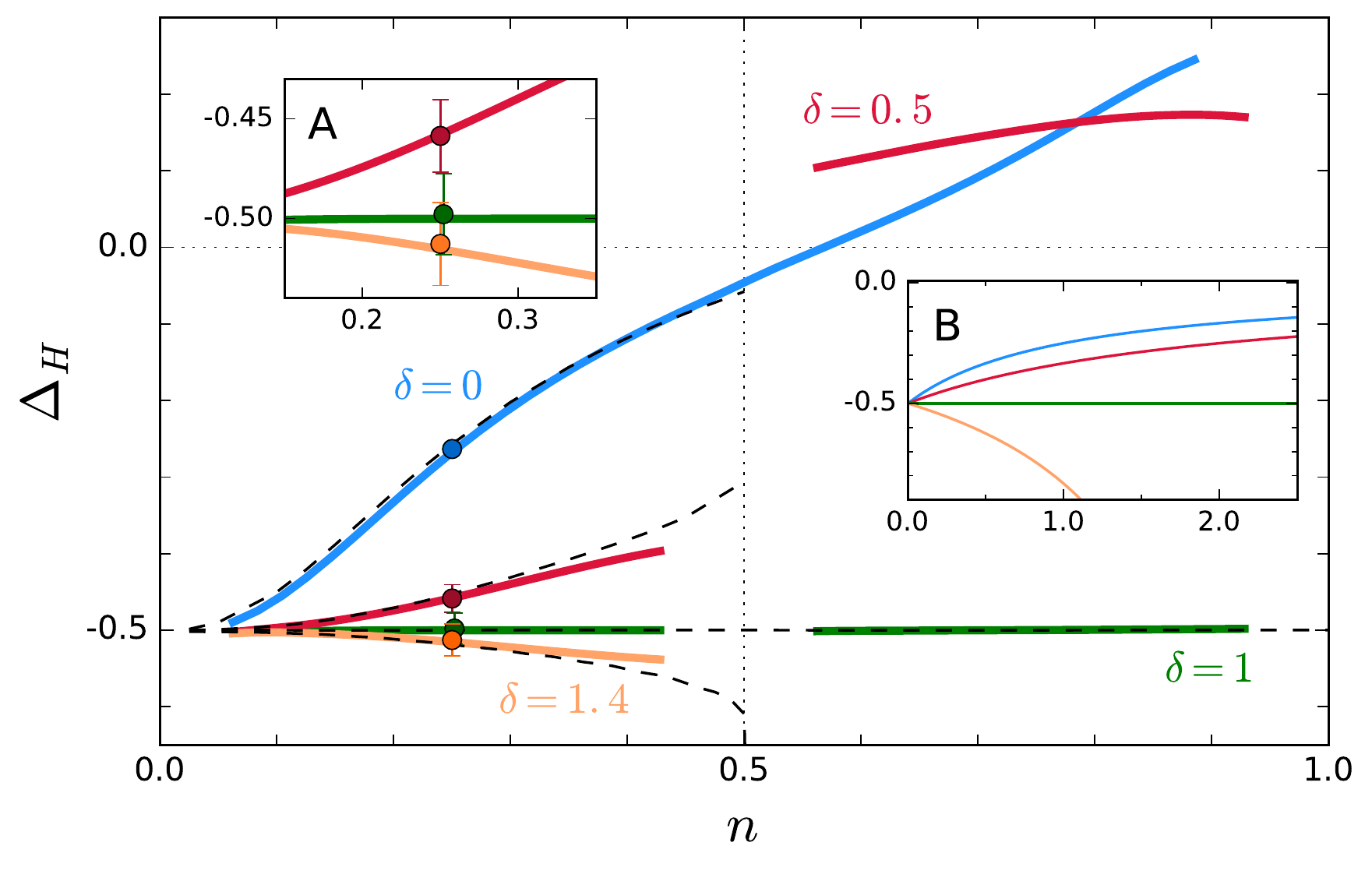}
\caption{
Hall effect out of $SU(2)$-symmetry for the Bose-Hubbard model. $\Delta_{\rm H}$ as function of the density $n$ for different values of interactions $\delta=U_{01}/U_{00} = 0$, $0.5$, $1.0$ and $1.4$ (extrapolated DMRG data for $L=24$, $28$ and $32$ rungs, $n_{max}=2$, $U_{00}=24t$, $t_y=t_x$). Dashed lines depict $\Delta_{\rm H}'$. Symbols with errorbars are time averages of the quench simulations for $10<\tau/t<20$ ($L=48$ rungs, $\Delta \mu=0.05t$, $\chi=0.01$).
Inset A shows a zoomed view of the plot. Inset B depicts the coherent state approximation Eq.~\eqref{eq:DHcoh} for $U_{00}=t_x=t_y$ and the same values of $\delta=0$, $0.5$, $1$ and $1.4$ (top to bottom).
}
\label{fig:3}
\end{figure}

Examples for the generally more complex dependence of $\Delta_{\rm H}$ on filling and interaction strengths are shown in Fig.~\ref{fig:3} for a strongly interacting $M=2$ Bose-Hubbard model for several ratios of on-rung and on-site interactions $\delta = U_{01}/U_{00}$. In the low filling regime $n \to 0$ we observe a good qualitative agreement with the mean field result of Eq.~\eqref{eq:DHcoh} (inset of Fig.~\ref{fig:3}).
Generally the Hall imbalance may unveil the presence of phase transitions and gaped phases, e.g. we observe a finite jump in $\Delta_{\rm H}$ close to the commensurate-incommensurate phase transition at half filling (compare Fig.~\ref{fig:3}, $\delta=0.5$). The more remarkable is the vanishing of these features at the exact $SU(2)$ point ($\delta=1$ curve in Fig.~\ref{fig:3}).

{While the mean-field description leading to Eq.~\eqref{eq:DHcoh} indicates that for $U_{00}<U_{01}$ the Hall imbalance vanishes with increasing density or interaction strength, in the strong coupling regime for $\delta\to 0$ (and $n>1/2$) we find an interesting sign change of the Hall imbalance. This property can be attributed to the restored particle hole symmetry in the as in hardcore boson limit which leads to the same sign change as discussed above for free fermions. We may also understand this as a precursor of the topological phase transition previously reported previously by Huber et al. \cite{Berg2015, Huber2011}.}

{For large fillings and $\delta>\delta_c$, for some critical $\delta_c=\delta_c(n, t_y/t_x)>1$, we observe a quantum phase transition to a biased ladder phase (BLP), where a majority of the particles accumulates on one leg of the ladder~\cite{supmat}, corresponding to a ferromagnetic state with fully polarized spin~\cite{Wei2014,Greschner2015,Greschner2016}. Due to the spontaneous breaking of $\mathbb{Z}_2$ symmetry in the thermodynamic limit this state exhibits $\la P_y \ra \neq 0$ at $\chi=0$ and, hence, we expect a diverging Hall imbalance~\cite{supmat}. This can be also seen in the classical model Eq.~\eqref{eq:DHcoh} in which for $n > t_y / (U_{01} - U_{00})$ the system becomes unstable and develops a spontaneous imbalance $\la P_y \ra > 0$ at vanishing field, resulting in a diverging Hall imbalance. 
Possibly such a giant Hall response can be observed within ferromagnetic alkali species such as $^{23}{\rm Na}$ or with the help of tuning of scattering lengths by means of optical Feshbach resonances~\cite{Pagano2014}.}

%%%%%%%%%%%%%%%%%%%%%%%%%%%%%%%%%%%%%%%%%%%%%%%%%%%%%%%%%%%%%%%%%
%%%%%%%%%%%%%%%%%%%%%%%%%%%%%%%%%%%%%%%%%%%%%%%%%%%%%%%%%%%%%%%%%
%%%%%%%%%%    OBC STORY
%%%%%%%%%%%%%%%%%%%%%%%%%%%%%%%%%%%%%%%%%%%%%%%%%%%%%%%%%%%%%%%%%
%%%%%%%%%%%%%%%%%%%%%%%%%%%%%%%%%%%%%%%%%%%%%%%%%%%%%%%%%%%%%%%%%

We conclude by discussing the particular case in which $[J_x,\mathcal H]=0$, valid {\it e.g.} for non-interacting particles, but also interacting fermions when projected on the lower bands. In this case, the off-diagonal matrix elements of $J_x$ vanish, leading to a compact simplified expression for $\Delta_{\rm H}$, $\Delta_{\rm H}' =\sum_{\alpha\neq 0} \frac{\la 0|P_y|\alpha\ra}{\mathcal E_0-\mathcal E_\alpha}  \frac{\la \alpha|\tilde{T}_x |0\ra}{\la 0| T_x |0\ra}$~\cite{supmat}.
For interacting systems, one generally finds  $\Delta_{\rm H}'\neq \Delta_{\rm H}$ (except for the interesting cases mentioned above), but the former is a remarkably good approximation in many cases, see Figs.~\ref{fig:2} and~\ref{fig:3}. It is also remarkable  that $\Delta'_{\rm H}$ can be efficiently evaluated with open-boundary conditions~(OBC) as well by means of DMRG calculations~\cite{supmat}. 

%%%%%%%%%%%%%%%%%%%%%%%%%%%%%%%%%%%%%%%%%%%%%%%%%%%%%%
%%%%%%%%%%%%%%%%%%%%%%%%%%%%%%%%%%%%%%%%%%%%%%%%%%%%%%
%%%%%%%   CONCLUSIONS
%%%%%%%%%%%%%%%%%%%%%%%%%%%%%%%%%%%%%%%%%%%%%%%%%%%%%%
%%%%%%%%%%%%%%%%%%%%%%%%%%%%%%%%%%%%%%%%%%%%%%%%%%%%%%

{\em Summarizing}, we have studied the  Hall response  of an interacting ladder. While generally the Hall effect strongly depends on the precise type and form of interactions, for certain single component states of $SU(M)$ symmetric models we observed the Hall imbalance $\Delta_{\rm H}$ to be independent from filling and interactions strength, corresponding to a universal $1/n$ behavior of the Hall constant $R_{\rm H}$.
In this work we have focused on the cases relevant or compatible with the experimental measurement procedures of Ref.~\cite{Genkina2018}, in general however, the reactive Hall coefficient may depend on the details of the measurement procedure~\cite{Filippone2018hall}. 
Further interesting extensions could include the role of interactions at finite field strengths where already in two-ladder systems a wealth of quantum phases and phenomena has been reported~\cite{Orignac2001,Carr2006,Dhar2012,Wei2014,Piraud2015,Greschner2015,Barbarino2016,Greschner2017,Petrescu2017}, which might exhibit unconventional Hall responses observable in current quantum gas experiments.

%%%%%%%%%%%%%%%%%%%%%%%%%%%%%%%%%%%%%%%%%%%%%%%%%%%%%%
%%%%%%%%%%%%%%%%%%%%%%%%%%%%%%%%%%%%%%%%%%%%%%%%%%%%%%
%%%%%%%   ACKNOWLEDGMENTS
%%%%%%%%%%%%%%%%%%%%%%%%%%%%%%%%%%%%%%%%%%%%%%%%%%%%%%
%%%%%%%%%%%%%%%%%%%%%%%%%%%%%%%%%%%%%%%%%%%%%%%%%%%%%%

\begin{acknowledgments}
We thank Assa Auerbach, Christophe Berthod,  Nigel Cooper, Temo Vekua, Ian Spielman and Leonardo Fallani for important discussion and in particular Charles E. Bardyn for insightful comments and remarks throughout the realization of this work. We also acknowledge support by the Swiss National Science Foundation under Division II.
Simulations were in parts carried out on the cluster system at the Leibniz University of Hannover, Germany and the Baobab cluster of the Unversity of Geneva, Switzerland.
\end{acknowledgments}

\bibliography{references}

\end{document}

\title{Supplemental Materials to ``Universal Hall Response in Synthetic Dimensions''}
\author{Sebastian Greschner}
\author{Michele Filippone}
\author{Thierry Giamarchi}
\affiliation{Department of Quantum Matter Physics, University of Geneva, 1211 Geneva, Switzerland}

\begin{abstract}
In the supplementary material we give details of the analytical calculations and numerical methods presented in the main text. We show additional examples for Hall imbalance in $M=2$, $3$ and $4$ leg ladders and compare the quench time evolution of the Hall imbalance in an interacting system to our results from DMRG calculations. We furthermore discuss the quantum phase transitions and observables observed for the interacting boson models with and without $SU(2)$ symmetry.
\end{abstract}
\date{\today}

\maketitle

%%%%%%%%%%%%%%%%%%%%%%%%%%%%%%%%%%%%%%%%%%%%%%%%%%%%%
%%%%%%%%%%%%%%%%%%%%%%%%%%%%%%%%%%%%%%%%%%%%%%%%%%%%%
%%%%%%%%  ANALYTICS
%%%%%%%%%%%%%%%%%%%%%%%%%%%%%%%%%%%%%%%%%%%%%%%%%%%%%
%%%%%%%%%%%%%%%%%%%%%%%%%%%%%%%%%%%%%%%%%%%%%%%%%%%%%
\section{Reactive Hall response}

In the following we present details on the derivation of the analytical expressions for $\Delta_H$ and $R_H$ presented in the main text. For clarity, we recall here the HH Hamiltonian with an ``on-leg gauge'' prescription for the magnetic field $\chi$
\begin{align}
H(\phi, \chi, E_y) &=H^x_{\rm kin}+H^y_{\rm kin}+H_{\rm int} + E_y P_y\nonumber	\\
H_{\rm int}^x&=  -t_x \sum_{j,m} \e^{\ii ((m-m_0)\chi + \phi/L)} a_{j,m}^\dagger a^{\phantom \dagger}_{j+1,m} + {\rm H.c.} \nonumber\\
H_{\rm kin}^y&=- t_y \sum_{j,m} a_{j,m+1}^\dagger a^{\phantom \dagger}_{j,\xi+1} + {\rm H.c.} \nonumber\\
H_{int}&=\sum_{j,m,m'}\frac{U_{m,m'}}2n_{j,m}n_{j,m'}
\label{eq:ham}
\end{align}
in which $j\in[1,L]$, $m\in[0,M-1]$ and we choose a symmetric gauge centered at  $m_0=(M-1)/2$.

\subsection{General approach}
We now follow the derivation of an expression for $R_H$ in terms of ground state energy derivatives by Prelov\v sek and collaborators presented in Ref.~\cite{Prelovsek1999,*Zotos2000} and generalize it to a formula for the Hall imbalance $\Delta_{\rm H}$ that we use in the main text. 
It is useful for the following discussion to consider the expansion of the ground state energy $\mathcal E_0(\phi,\chi,E_y)$ to third order in $\phi,\chi,E_y$ close to zero:
\begin{equation}\label{eq:expansion}
\begin{split}
\mathcal E_0(\phi,\chi,E_y)=&\mathcal E_0(0,0,0)+\frac{\phi^2}2 \frac{\partial^2 \mathcal E_0}{\partial\phi^2}+\frac{\chi^2}2 \frac{\partial^2 \mathcal E_0}{\partial\chi^2}\\&+\frac{E_y^2}2 \frac{\partial^2 \mathcal E_0}{\partial E_y^2}+\phi\chi E_y\frac{\partial^3\mathcal E_0}{\partial\phi\partial \chi\partial E_y}+\ldots
\end{split}
\end{equation}
in which we discarded the terms which vanish because of the symmetry of the problem. 
Current and polarization are both given by derivatives of the energy, namely
\begin{align}
\langle J_x\rangle&=L\frac{\partial\mathcal E_0}{\partial \phi}\,,
&
\langle P_y \rangle&=\frac{\partial\mathcal E_0}{\partial E_y}\,.
\end{align}
Close to $(\phi,\chi,E_y)=(0,0,0)$ the expansion ~\eqref{eq:expansion} leads to
\begin{align}\label{eq:jxpy}
\langle J_x\rangle&=L\phi\frac{\partial^2 \mathcal E_0}{\partial\phi^2}\,, & \langle P_y\rangle&=E_y\frac{\partial^2 \mathcal E_0}{\partial E_y^2}+\phi\chi\frac{\partial^3\mathcal E_0}{\partial\phi\partial \chi\partial E_y}\,.
\end{align}
For, the measurement/calculation of the Hall imbalance $\Delta_{\rm H}$, Eq.~\eqdh in the main text, no transverse field is required ($E_y=0$), thus the expression for $\Delta_{\rm H}$ reads
\begin{equation}
\Delta_{\rm H}=\frac1L\left.\frac{\frac{\partial\mathcal E_0}{\partial\phi\partial\chi\partial E_y}}{\frac{\partial^2\mathcal E_0}{\partial \phi^2}}\right|_{\phi,\chi,E_y=0}\,,
\end{equation} 
which is also given in the main text.

The calculation of $R_{\rm H}$ requires to derive the electric field $E_y$, such that $\langle P_y\rangle=0$. Equation~\eqref{eq:jxpy} leads then to the condition
\begin{equation}
E_y=-\phi\chi\frac{\frac{\partial^3\mathcal E_0}{\partial\phi\partial \chi\partial E_y}}{\frac{\partial^2 \mathcal E_0}{\partial E_y^2}}\,,
\end{equation}
leading to  the expression for $R_{\rm H}$
\begin{equation}
R_{\rm H}=\left.\frac{-L\Delta_{\rm H}}{\frac{\partial^2\mathcal E_0}{\partial E_y^2}}\right|_{\phi,\chi,E_y=0}\,,
\end{equation} 
which is also given in the main text.

\subsection{Non-interacting particles}
In this section, we carry out the explicit calculation of $\Delta_{\rm H}$ and $R_{\rm H}$ leading to Eq.~\eqDeltaHfreefermions in the main text. The expansion~\eqref{eq:expansion} can be derived by second order perturbation theory of the Harper-Hofstadter (HH) Hamiltonian in $\chi$ and $E_y$.
It is insightful to work in the basis $\{|k,p\rangle\}$ which diagonalizes the HH Hamiltonian for $(\chi,E_y)=0$, namely
\begin{equation}\label{eq:basis}
\langle j,m|k,p\rangle=\sqrt{\frac{2}{L(M+1)}}e^{ikx}\sin \big[n_p(m-m_0+1)\big]\,, 
\end{equation}
with $k=2\pi n_k/L$ ($n_k\in [0,L-1]$) and $n_p=\pi (p+1)/(M+1)$ ($p\in[0,M-1]$). For our purposes, it is enough to expand the single-particle Hamiltonian linearly in $\chi$, such that the  Hamiltonian can be cast in the form $H=H_0+V$, with $H_0=H(\phi,\chi=0)$ and
\begin{align}
V&=\chi \tilde J_x+E_yP_y\,,\\
\tilde J_x&=-i t_x\sum_{j,m}(m-m_0)\Big[e^{i\phi/L}a^\dagger_{j,m}a_{j+1,m}+\mbox{H.c.}\Big]\,.
\end{align}
Switching to the diagonal basis~\eqref{eq:basis}, one finds 
\begin{align}
H_0&=\sum_{kp}\varepsilon^{(0)}(k,p)a^\dagger_{k,p}a_{k,p}\,,\\
V&=\sum_{kpp'}\left[\chi \,v_x\left(k+\frac\phi L\right)+E_y\right]C_{p,p'}a^\dagger_{k,p}a_{k,p'}\,,
\end{align}
in which $\varepsilon^{(0)}(k,p)a^\dagger_{k,p}a_{k,p}=\varepsilon_x(k+\phi/ L)+\varepsilon_y(p)$, $v_x(k)=\partial_k\varepsilon_x(k)$ is the velocity along $x$ of a state of wave-vector $k$  and 
\begin{equation}\label{eq:cpp}
\begin{split}
C_{p,p'}&=\frac2{M+1}\sum_{m=0}^{M-1}(m-m_0)\times\\&\sin[n_p(m-m_0+1)]\sin[n_{p'}(m-m_0+1)]\,.
\end{split}
\end{equation}
The expansion of the single-particle energies in $V$ reads $\varepsilon(k,p)=\varepsilon^{(0)}(k,p)+\varepsilon^{(2)}(p)$, in which the leading contribution to the single particle energies  is of second order in $V$, $k$-independent  and reads 
\begin{equation}\label{eq:2ndorder}
\begin{split}
\varepsilon^{(2)}(p)&=\mathcal I_p(M,t_y)\left[\chi \,v_x\left(k+\frac\phi L\right)+E_y\right]^2\,,\\
\mathcal I_p(M,t_y)&=\sum_{p'\neq p}
\frac{C_{p,p'}^2}{\varepsilon_y(p)-\varepsilon_y(p')}\,,
\end{split}
\end{equation}
in which we highlighted the dependence of the factor $\mathcal I_p$ on the lattice parameter $t_y$ and width $M$. Equation~\eqref{eq:2ndorder} can be further expanded in $\phi/L$
\begin{equation}
\varepsilon^{(2)}(p)=\mathcal I_p(M,t_y)\left[\chi \,v_x(k)+\chi\frac\phi Lm_x^{-1}(k)+E_y\right]^2
\end{equation}
in which the effective inverse mass $m_x^{-1}(k)=\partial_kv_x(k)$ has been introduced. A term of order $(\phi/L)^2$ comes out from the expansion of the unperturbed energies as well $\varepsilon_x(k+\phi/L)=\varepsilon_x(k)+\phi v_x(k)/L+\phi^2 m_x^{-1}(k)/2L^2$. In the non-interacting case, the ground state energy of the system is the sum of the single particle energies $\mathcal E_0=\sum_{k,p}f_{k,p}\varepsilon(k,p)$, in which $f_{k,p}$ is the single particle distribution function describing the ground state. The derivatives of interest for the calculation of $\Delta_{\rm H}$ and $R_{\rm H}$ read
\begin{equation}\label{eq:dergen}
\begin{aligned}
\frac{\partial^2 \mathcal E_0}{\partial \phi^2}=&\frac 1{L^2}\sum_{k,p} f_{k,p} \,m_x^{-1}(0)\,,\quad\frac{\partial^2 \mathcal E_0}{\partial E_y^2}=\sum_{k,p}f_{k,p}\mathcal I_p(M,t_y)\,,\\
&\frac{\partial^3 \mathcal E_0}{\partial \phi\partial\chi\partial E_y}=\frac 1L\sum_{k,p}f_{k,p} \,m_x^{-1}(k)\mathcal I_p(M,t_y)\,.
\end{aligned}
\end{equation} 
Notice that $f_{k,p}$ in general cannot be factorized as separate functions of exclusively $k$ and $p$. This will appear in the following calculations for free bosons and fermions.  

\begin{figure}[tb]
\centering
\includegraphics[width=0.49\linewidth]{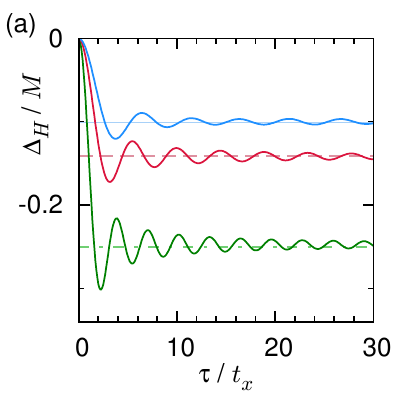}
\includegraphics[height=0.49\linewidth]{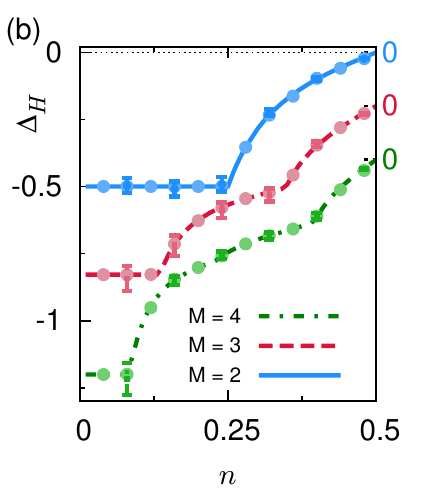}
\caption{
(a) Dynamics of the Hall imbalance $\Delta_H (\tau)$ during a quenched tilt for free fermions for (bottom to top) $M=2$, $3$ and $4$ legs ($n=N/L=1/5$, $\chi=0.01$, $\Delta\mu=0.01$, exact diagonalization~(ED), $L=160$ rungs). 
(b) $\Delta_H$ for free fermions on $M$-leg ladders from various approaches. Solid lines correspond to Eq.~\eqDeltaHSUM of the main text, data points depict $\Delta_H'$ to open-boundary conditions~(OBC) (ED, $L=80$ rungs) and points with error bars time averages of the quench dynamics (curves for $M=4$ and $M=3$ have been shifted down for clarity. Note, that for all cases $\Delta_H=0$ for $n=0.5$). 
}
\label{fig:S1}
\end{figure}

\subsubsection{Bosons}
In the ground state of non-interacting bosons, all $N$ particles collapse on the lowest single-particle energy state, which, in the usual case and in particular for the HH model, corresponds to be at $k=p=0$. The derivatives of interest for $\Delta_{\rm H}$ and $R_{\rm H}$ read
\begin{equation}\label{eq:derbosons}
\begin{aligned}
\frac{\partial^2 \mathcal E_0}{\partial \phi^2}=&\frac {N}{L^2}  \,m_x^{-1}(p)\,,\qquad\frac{\partial^2 \mathcal E_0}{\partial E_y^2}=N\mathcal I_0(M,t_y)\,,\\
&\frac{\partial^3 \mathcal E_0}{\partial \phi\partial\chi\partial E_y}=\frac NL \,m_x^{-1}(0)\mathcal I_0(M,t_y)\,.
\end{aligned}
\end{equation} 
One thus directly finds Eq.~\eqDeltaHSUM in the main text.

\subsubsection{Fermions}
The ground state of non interacting fermions is given by  $f_{k,p}=\theta[\mu-\varepsilon(k,p)]$ in which $\mu$ is the Fermi energy. We neglect here important parity effects~\cite{filippone2018controlled}. We also switch to the continuum limit for the wave-vector $k$ to perform the sums in Eq.~\eqref{eq:dergen}. The spectrum $\varepsilon(k,p)$ is composed of $M$ non-crossing bands labeled by the index $p$. The Fermi Energy $\mu$ crosses a limited amount of bands $p\leq P$ at different Fermi wave-vectors $k_{F,p}$, see also Fig.~1 of the main text.  Thus, the fermionic equivalent of Eq.~\eqref{eq:derbosons} reads
\begin{equation}\label{eq:derfermions}
\begin{aligned}
\frac{\partial^2 \mathcal E_0}{\partial \phi^2}=& \sum_{p<P} \frac{v_{F,p}}{\pi L}\,,\quad\frac{\partial^2 \mathcal E_0}{\partial E_y^2}=\sum_{p<P}N_p\mathcal I_p(M,t_y)\,,\\
&\frac{\partial^3 \mathcal E_0}{\partial \phi\partial\chi\partial E_y}=\sum_{p<P}\frac{v_{F,p}}\pi\mathcal I_p(M,t_y)\,.
\end{aligned}
\end{equation} 
in which we stress that $N_{p}=Lk_{F,p}/\pi$ is the number of particles occupying band $p$ and $v_{F,p}=v_x(k_{F,p})$ its Fermi velocity. This leads to Eq.~\eqDeltaHfreefermions in the main text and, when only the lower $p=0$ band is occupied, one recovers Eq.~\eqDeltaHSUM. In Fig.~\ref{fig:S1} we show several examples of $\Delta_H$ for $M=2$, $3$ and $4$ leg ladders.

%%%%%%%%%%%%%%%%%%%%%%%%%%%%%%%%%%%%%%%%%%%%%%%%%%%%%%%
%%%%%%%%%%%%%%%%%%%%%%%%%%%%%%%%%%%%%%%%%%%%%%%%%%%%%%%
%%%%%%% PERTURBATION THEORY MANY-BODY
%%%%%%%%%%%%%%%%%%%%%%%%%%%%%%%%%%%%%%%%%%%%%%%%%%%%%%%
%%%%%%%%%%%%%%%%%%%%%%%%%%%%%%%%%%%%%%%%%%%%%%%%%%%%%%%
\section{Hall Response for Interacting particles}
We generalize now to the interacting/many-body setting.  In order to calculate the weak field Hall effect, we will assume that the fluxes $\phi$ and $\chi$ are small. In the case of periodic boundary conditions (PBC) one has to care about the quantization of $\chi=2\pi/L\cdot \#$ (with some integer $\#$) and important parity effects~\cite{filippone2018controlled}. Let us ignore this subtlety for the moment and expand the Hamiltonian~\eqref{eq:ham} up to second order in $\chi$:
\begin{equation}
\begin{aligned}
H = &H_0(\phi, \chi=0)+ \chi \tilde{J} - \frac{\chi^2}{2} \dbtilde{T} + \cdots 
\end{aligned}
\end{equation}
in which we introduced the following anti-symmetric current ($\tilde J_x$) and kinetic ($\tilde T_x$) operators
\begin{align}
\label{eq:tildejx}\tilde{J_x}(\phi) &= -t_x\sum_{j,m} \ii (m-m_0) \,e^{\ii\frac\phi L} a_{j,m}^\dagger a^{\phantom \dagger}_{j+1,m} + {\rm H.c.}\\
\tilde{T_x}(\phi) &= -t_x\sum_{j,m} (m-m_0) \,e^{\ii \frac\phi L} a_{j,m}^\dagger a^{\phantom \dagger}_{j+1,m} + {\rm H.c.}
\end{align}
We want to calculate the Hall imbalance $\Delta_{\rm H}$, see Eq.~\eqdh in the main text, which requires the averages $\la P_y\ra$ and $\la J_{x}\ra$.
For $\chi=0$ and $\phi\neq0$, due to the reflection symmetry of the Hamiltonian $H$ and anti-symmetry of the polarization $P_y$ with respect to reflection in space in $y$ direction, the polarization vanishes. The leading contribution to $\la P_y\ra$ is linear in $\chi$ and Eq.~\eqpertDeltan in the main text is readily found after expanding the ground state $|0\ra$ to linear order in $\chi$
\begin{equation}\label{eq:perteigenstates}
|0\ra_\chi=|0\ra_\phi+\sum_{\alpha\neq0}\frac{~_\phi\la\alpha|\tilde J_x(\phi)|0\ra_\phi}{\mathcal E_0-\mathcal E_\alpha}|\alpha\ra_\phi
\end{equation}
in which all eigenstates $\{|\alpha\ra_{\phi}\}$ are eigenstates of the Hamiltonian~\eqref{eq:ham} for $\chi=0$ but finite $\phi$. We report Eq.\eqpertDeltan for clarity 
\begin{equation}\label{eq:pertDeltansup}
\begin{split}
~_\phi\la 0| P_y|0 \ra_\phi =& \chi \sum_{\alpha\neq 0}\frac{~_\phi\la 0| P_y |\alpha \ra_\phi~_\phi \la \alpha| \tilde{J}_x(\phi) |0 \ra_\phi+\mbox{c.c.}}{\mathcal E_0 -\mathcal  E_\alpha}
\end{split}\,.
\end{equation}
The expansion of the above expression to leading order in $\phi$ leads to 
\begin{widetext}
\begin{equation}\label{eq:polhorror}
\begin{split}
\la0| P_y|0 \ra =& - 2\frac{\phi\chi}L\sum_{\alpha\neq 0}\frac{\la 0| P_y |\alpha \ra \la \alpha| \tilde T_x |0 \ra}{\mathcal E_0 - \mathcal E_\alpha} +\frac{\chi\phi}L
 \sum_{\alpha,\alpha'\neq 0}\left[\frac{J_x^{0\alpha}P_y^{\alpha'\alpha}\tilde J_x^{\alpha0} + P_y^{0\alpha}\tilde J_x^{\alpha\alpha'}J_x^{\alpha'0}}{(\mathcal E_0 - \mathcal E_\alpha)(\mathcal E_0 -\mathcal  E_{\alpha'})} +\mbox{c.c}\right]\\
&+\frac{\chi\phi}L
\sum_{\alpha\neq\alpha'}\left[\frac{P_y^{0\alpha'}\tilde J_x^{\alpha'\alpha}J_x^{\alpha 0}+P_y^{0\alpha}\tilde J_x^{\alpha\alpha'}J_x^{\alpha'0}}{(\mathcal E_0 - \mathcal E_\alpha)(\mathcal E_\alpha -\mathcal  E_{\alpha'})} +\mbox{c.c}\right]
\end{split} 
\end{equation}
\end{widetext}
in which we use the shorthand notation $\mathcal O^{\alpha\beta}=\la\alpha|\mathcal O|\beta\ra$ for the matrix elements, applied $\tilde J_x(\phi)=\tilde J_x-\phi\tilde T_x/L$ (we also use the shorthand notation $\tilde J_x=\tilde J_x(0)$) from Eq.~\eqref{eq:tildejx} and an expansion analog to Eq.~\eqref{eq:perteigenstates} for the eigenstates in $\phi$ which couples to the symmetric current operator $J_x$.

The well-known expression for the current in terms of Drude weight $D$ is found by second order expansion in $\phi$ and $\chi=0$~\cite{kohn64,shastry90,millis90} 
\begin{align}\label{eq:D}
\begin{split}
\la J_{x}\ra &= L\frac{\partial \mathcal E_0}{\partial \phi}  = \phi \cdot D\,, \\
D &= - \frac{\la T_x \ra }L+ \frac 2L \sum_{\alpha\neq 0}\frac{|\la 0| J_x |\alpha \ra|^2}{E_0 - E_\alpha}\,,
\end{split}
\end{align}
in which $T_x=-t_x\sum_{j,m}a^\dagger_{j,m}a_{j+1,m}+\mbox{h.c.}$ is the kinetic energy operator. The ratio of Eqs.~\eqref{eq:polhorror} and~\eqref{eq:D} leads to the Hall imbalance $\Delta_{\rm H}$. To derive the Hall constant $R_{\rm H}$ one needs the transverse compressibility $\partial^2 \mathcal E_0/\partial E_y^2$ at $\chi,\phi=0$, which is also readily derived in perturbation theory 
\begin{equation}\label{eq:compressibility}
\frac{\partial^2 \mathcal E_0}{\partial E_y^2}=\sum_{\alpha\neq0}\frac{|\la \alpha|P_y|0\ra|^2}{\mathcal E_0-\mathcal E_\alpha}\,.
\end{equation}

%%%%%%%%%%%%%%%%%%%%%%%%%%%%%%%%%%%%%%%%%%%%%%%%%%%%
%%%%%%%%%%%%%%%%%%%%%%%%%%%%%%%%%%%%%%%%%%%%%%%%%%%%
%%%%%%%%%%% UNIVERSAL REGIME ANALYTICS
%%%%%%%%%%%%%%%%%%%%%%%%%%%%%%%%%%%%%%%%%%%%%%%%%%%%
%%%%%%%%%%%%%%%%%%%%%%%%%%%%%%%%%%%%%%%%%%%%%%%%%%%%

\begin{figure}[tb]
\centering
\includegraphics[width=1\linewidth]{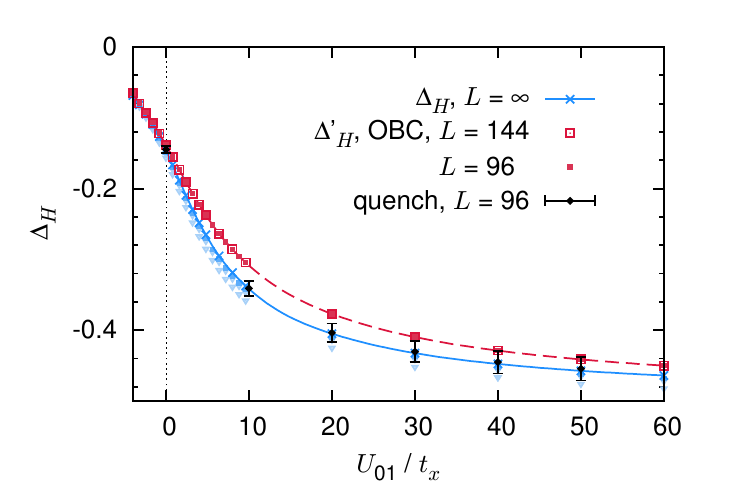}
\caption{
Comparison of methods for the calculation of $\Delta_{\rm H}$ and $\Delta_{\rm H}'$ for hardcore bosons $n_{max}=1$ as function of $U_{01}$ ($t_y=t_x$). PBC results (blue curves and symbols) have been extrapolated to the thermodynamic limit using a second order polynomial in $1/L^2$. The results coincide precisely with the time-average of a quenched tilt simulation (symbols with errorbars) for $L=96$ rungs. The red curve and symbols correspond to $\Delta_{\rm H}'$ as derived for $L=96$ and $L=144$ rungs systems and deviate considerably from $\Delta_{\rm H}$ for finite $U$.}
\label{fig:S2}
\end{figure}

\subsubsection{Hall imbalance and Hall constant in the universal regime}
As mentioned in the main  text, the universality of the Hall response can be understood by first switching to the basis diagonalizing $H_{\rm kin}^y$ in Eq.~\eqref{eq:ham}. This basis is given by the transformation
\begin{equation}
a_{j,m}=\sum_{p}\sqrt{\frac2{M+1}}\sin[n_p(m-m_0+1)]a_{j,p}
\end{equation}
with the $p$ labels as in Eq.~\eqref{eq:basis}. In this basis we have 
\begin{align}
P_y&=\sum_{j,p,p'}C_{p,p'}a^\dagger_{j,p}a_{j,p'}\,, \\
\tilde J_x(\phi)&=-t_x\sum_{j,p,p'}\ii C_{p,p'}e^{i\phi/L}a^\dagger_{j,p}a_{j,p'}+\mbox{H.c.}
\end{align}
with the coefficients $C_{p,p'}$ given in Eq.~\eqref{eq:cpp}. If we consider   stabilized fully polarized (FP) ground states in the $p=0$ component, the above operators acting on the ground state $|0\ra$ have finite matrix elements only with states $|\alpha_p\ra$ with one single excitations in $p>0$ components. As a consequence, from Eq.~\eqref{eq:pertDeltansup} (valid for $\phi\neq0$)
\begin{equation}
\begin{split}
~_\phi\la 0|P_y|0\ra_\phi=\chi& \sum_{p,j,j'}\frac{C_{0,p}^2}{\varepsilon_y(0)-\varepsilon_y(p)}\times\\
&(-t_x) \la 0|\ii e^{\ii\phi/L}a^\dagger_{j,0}a_{j+1,0}+\mbox{h.c.}|0\ra\\
=\chi&\mathcal I_0~_\phi\la0|J_x(\phi)|0\ra_\phi\,,
\end{split}
\end{equation}
with the band coefficient $\mathcal I_p$ given in Eq.~\eqref{eq:2ndorder}. We have here used $\mathcal E_{\alpha_p} - \mathcal E_0 = \epsilon_y(p) - \epsilon_y(0)$ as discussed in the main text. This result leads to the universal expression of $\Delta_{\rm H}$, Eq.~\eqDeltaHSUM in the main text.

The universality of the Hall constant $R_{\rm H}$ follows from an analogous calculation for the compressibility~\eqref{eq:compressibility}
\begin{equation}
\frac{\partial^2 \mathcal E_0}{\partial E_y^2}=n\mathcal I_0\,,
\end{equation}
which leads to Eq.~\eqDeltaHSUM.

We stress here that this derivation applies for both fermionic and bosonic $SU(M)$ symmetric FP ground states in the presence of interactions.  

%%%%%%%%%%%%%%%%%%%%%%%%%%%%%%%%%%%%%%%%%%%%%%%%%%%%%
%%%%%%%%%%%%%%%%%%%%%%%%%%%%%%%%%%%%%%%%%%%%%%%%%%%%%
%%%%%%%%%%%%%   CURRENT CONSERVATION 
%%%%%%%%%%%%%%%%%%%%%%%%%%%%%%%%%%%%%%%%%%%%%%%%%%%%%
%%%%%%%%%%%%%%%%%%%%%%%%%%%%%%%%%%%%%%%%%%%%%%%%%%%%%

\subsubsection{Current conservation}
For models in which the current commutes with the Hamiltonian: $[J_x,H]=0$, which is for instance the case of non-interacting particles, eigenstates are labeled by the current operator and $\la \alpha|J_x|\beta\ra$ for $|\alpha\ra\neq|\beta\ra$. The low flux expansions~\eqref{eq:polhorror} and~\eqref{eq:D} are considerably simplified 
\begin{align}
\la P_y\ra&=-2\frac{\chi\phi}L\sum_{\alpha\neq0}\frac{\la0|P_y|\alpha\ra\la\alpha|\tilde T_x|0\ra}{\mathcal E_0-\mathcal E_\alpha}\,,\\
D&=-\frac{\la T_x\ra}L\,,
\end{align}
and the Hall imbalance reads 
\begin{equation}\label{eq:deltahprime}
\Delta_{\rm H}'= \sum_{\alpha\neq 0} \frac{\la 0|P_y|\alpha\ra}{E_0-E_{\alpha}}  \frac{\la \alpha|\tilde{T}_x |0\ra}{\la 0|T_x |0\ra}\,.
\end{equation}

%%%%%%%%%%%%%%%%%%%%%%%%%%%%%%%%%%%%%%%%%%%%%%%%%%%%%
%%%%%%%%%%%%%%%%%%%%%%%%%%%%%%%%%%%%%%%%%%%%%%%%%%%%%
%%%%%%%%%%%%%   HALL WITH OBC  
%%%%%%%%%%%%%%%%%%%%%%%%%%%%%%%%%%%%%%%%%%%%%%%%%%%%%
%%%%%%%%%%%%%%%%%%%%%%%%%%%%%%%%%%%%%%%%%%%%%%%%%%%%%

\subsubsection{Derivation of the Hall response with open boundary conditions}
The kinetic operators $T_x$ and $\tilde T_x$, which are present in Eq.~\eqref{eq:deltahprime}, can have finite averages with open boundary conditions (OBC) as well. One could then  wonder whether the Hall responses of interest in this work could be calculated in OBC settings as well. In this section we show that Eq.~\eqref{eq:deltahprime} could be directly calculated with open boundary conditions (OBC), what is extremely practical to perform DMRG calculations. One should nevertheless keep in mind that Eq.~\eqref{eq:deltahprime} is an approximation, but extremely efficient in the regimes of interest, as we are going to show in the following.

Consider the following Hamiltonian with OBC 
\begin{equation}\label{eq:hamprime}
H'=H(\phi=\chi=0)-\chi \tilde T_x\,,
\end{equation}
which is inspired by the leading contribution proportional to $\chi$ in the expansion of the Hamiltonian~\eqref{eq:ham}. The Hamiltonian~\eqref{eq:hamprime} explicitly breaks the reflection symmetry  around $m_0$ in the y-direction, leading to finite $\la P_y\ra\neq0$, namely 
\begin{align}
\la P_y\ra&=-2\frac{\chi}L\sum_{\alpha\neq0}\frac{\la0|P_y|\alpha\ra\la\alpha|\tilde T_x|0\ra}{\mathcal E_0-\mathcal E_\alpha}\,,
\end{align}
while also the Drude weight can be computed with OBC as $D=-\la T_x\ra/L$. It is important to note, that for OBC both Eq.~\eqref{eq:D} and Eq.~\eqref{eq:polhorror} vanish identically as current and polarization  can be expressed as energy derivatives wrt. $\phi$, and for OBC a trivial gauge transformation eliminate the uniform ``twist'' $\phi$ from the Hamiltonian.

%%%%%%%%%%%%%%%%%%%%%%%%%%%%%%%%%%%%%%%%%%%%%%%%%%%%%
%%%%%%%%%%%%%%%%%%%%%%%%%%%%%%%%%%%%%%%%%%%%%%%%%%%%%
%%%%%%%%%% BOSON MEAN-FIELD 
%%%%%%%%%%%%%%%%%%%%%%%%%%%%%%%%%%%%%%%%%%%%%%%%%%%%%
%%%%%%%%%%%%%%%%%%%%%%%%%%%%%%%%%%%%%%%%%%%%%%%%%%%%%

\section{Mean-field derivation of the Hall effect in the Bose-Hubbard model}
We derive here Eq.~\eqDHcoh in the main text. We consider the bosonic version of the model~\eqref{eq:ham} in the large density $n$ limit and approximate $a_{j,m}=\sqrt{n+(m-m_0)\delta n} \cdot \e^{\ii \varphi}$ and then find the $\delta n$ and $\varphi$ which minimizes the energy. For $M=2$, we have 
\begin{equation}
\begin{split}
H_{\rm kin}^x&\rightarrow-2t_xL\sum_{\sigma=\pm}\cos\left(\frac\phi L+\sigma\frac\chi2\right)\left(n+\sigma\frac{\delta n}2\right)\\
H_{\rm kin}^y&\rightarrow-2t_yL\sqrt{n^2-\frac{\delta n^2}4}\\
H_{\rm int}&=U_{00}L\left(n^2+\frac{\delta n^2}{4}\right)+U_{01}L\left(n^2-\frac{\delta n^2}4\right)
\end{split}
\end{equation}
Minimization of the energy with respect to $\delta n$ leads to the condition
\begin{equation}
\delta n=\frac{-4nt_x\sin(\phi/L)\sin(\chi/2)}{t_y+n(U_{00}-U_{01})}
\end{equation}
while the total current reads 
\begin{equation}
\la J_x\ra=L\partial_\phi \mathcal E_0=4t_xn L\sin(\phi/L)\cos(\chi/2)\,.
\end{equation}
Considering that the polarization reads $\la P_y\ra=L\delta n$, Eq.~\eqDHcoh is found in the $\phi,\chi\rightarrow 0$ limit. 

%%%%%%%%%%%%%%%%%%%%%%%%%%%%%%%%%%%%%%%%%%%%%%%%%%%%%
%%%%%%%%%%%%%%%%%%%%%%%%%%%%%%%%%%%%%%%%%%%%%%%%%%%%%
%%%%%%%%%% NUMERICAL METHODS 
%%%%%%%%%%%%%%%%%%%%%%%%%%%%%%%%%%%%%%%%%%%%%%%%%%%%%
%%%%%%%%%%%%%%%%%%%%%%%%%%%%%%%%%%%%%%%%%%%%%%%%%%%%%

\section{Numerical Methods}

\begin{figure}[tb]
\centering
\includegraphics[width=1\linewidth]{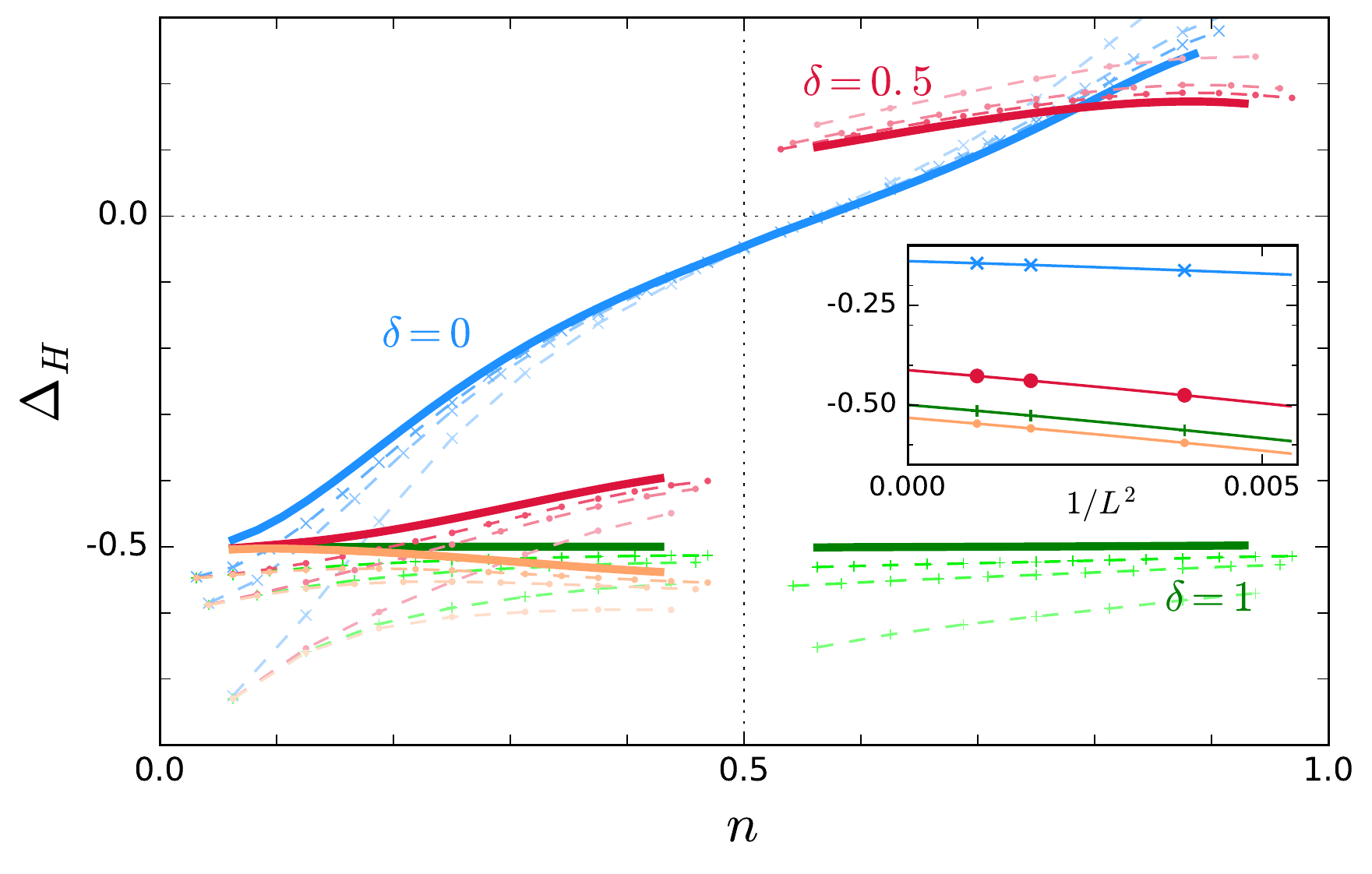}
\caption{Finite size scaling data of Fig.~3 of the main text, $\Delta_{\rm H}$ as function of the density $n$ (softcore bosons, $n_{max}=3$, $U_{00}=24t$). The dashed lines with the symbols indicate the PBC results for $L=16$, $24$ and $32$ rungs for different values of $\delta$. The inset shows an example of the $1/L^2$ scaling with the system size and the corresponding extrapolation for density $n=1/4$.}
\label{fig:S3}
\end{figure}

\begin{figure}[tb]
\centering
\includegraphics[width=1\linewidth]{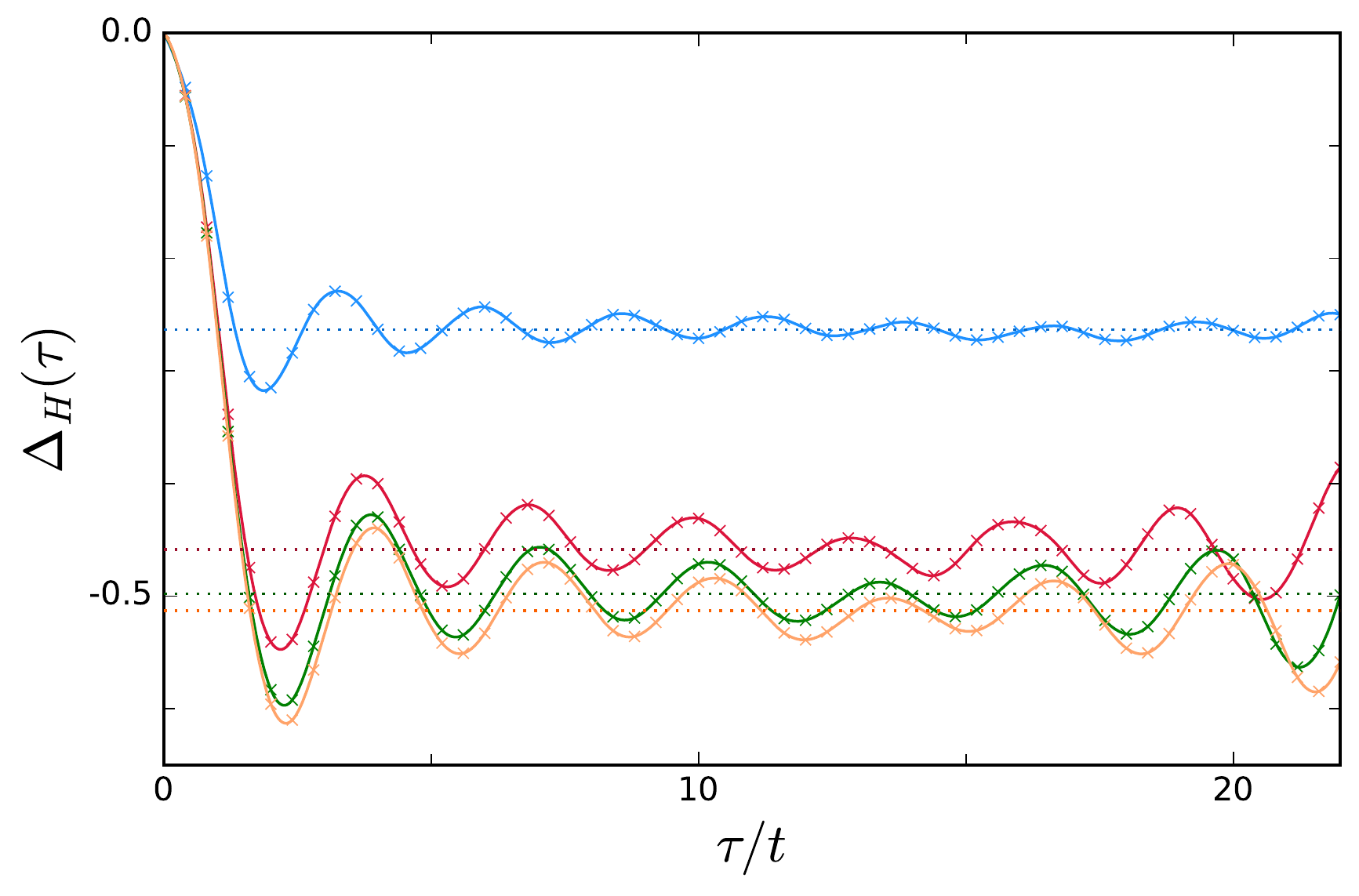}
\caption{Quench dynamics of the Hall imbalance $\Delta_{\rm H} (\tau)$ (symbols) for a tilt of $M=2$ softcore bosons as shown in Fig. 3 of the main text ($n=N/L=0.25$, $\chi=0.01$, $\Delta\mu=0.05t$, $L=48$ rungs, $n_{max}=3$). Dotted horizontal line shows the time average of $10<\tau / t<20$.}
\label{fig:S4}
\end{figure}

In the main text we introduce several methods and measurables for the Hall effect, which we will discuss in the following for several examples. A comparison of the methods is shown in Figs.~\ref{fig:S1} and ~\ref{fig:S2} for the example of free fermions and hardcore bosons ($U_{mm} \to \infty$, $U_{01}=0$) at quarter filling ($n=0.25$).

We directly simulate the system with periodic boundary conditions~(PBC) using DMRG for small system sizes up to $L=48$ rungs. In this case special care has to be taken for the choice of fluxes $\chi$ and $\phi$ to avoid parity effects~\cite{filippone2018controlled}. In particular for the bosonic system we choose $\chi=4\pi/L$ for systems with an even number of particles and some small value $\phi\sim 0.01$. We extrapolate the data to the thermodynamic limit with a polynomial in $1/L^2$ - examples are shown e.g. in Fig.~\ref{fig:S3} for a different example.

As shown in Fig.~\ref{fig:S2} the extrapolated results coincide precisely with data obtained form time dependent DMRG (tDMRG) simulations~\cite{Schollwoeck2011} for a quenched tilt potential as discussed in the main text. At $\tau=0$ the system $H(\chi,\phi=0)$ is initialized in its ground state(DMRG simulation with $D=400$). Subsequently the system evolves under the presence of a tilted potential $\Delta\mu \sum_{ij} i n_{ij}$ (tDMRG simulations with $\Delta\mu=0.05 t$, keeping $D<1000$). We extract the current $J_{x}$ and polarization on the central rung of the system and observe that the ratio $\Delta_{\rm H} (\tau) $ oscillates around some finite constant values. 
Time averaging over some intermediate range, such as $10 <\tau /t_x < 20$, we may obtain the average value with some variance shown as points with errorbars in Fig.~\ref{fig:S2}. 

The parameter $\Delta_{\rm H}'$ is evaluated within open boundary conditions~(OBC) for fermions and bosons on $L\leq 160$ rungs and up to $M=4$ leg ladders using DMRG. We typically we keep up to $D=1000$ matrix states. For calculations of $\Delta_{\rm H}'$ from a finite system, we calculate the ground-state for a small value of $\chi \ll 1$, typically $\chi \sim 0.01$ or $0.001$, and evaluate $\la P_y \ra$ and $\la J_{x}\ra$ for this state. Comparison to simulations with a smaller $\chi$ show the convergence of this method within the required margin of error.

\begin{figure}[tb]
\centering
\includegraphics[width=1\linewidth]{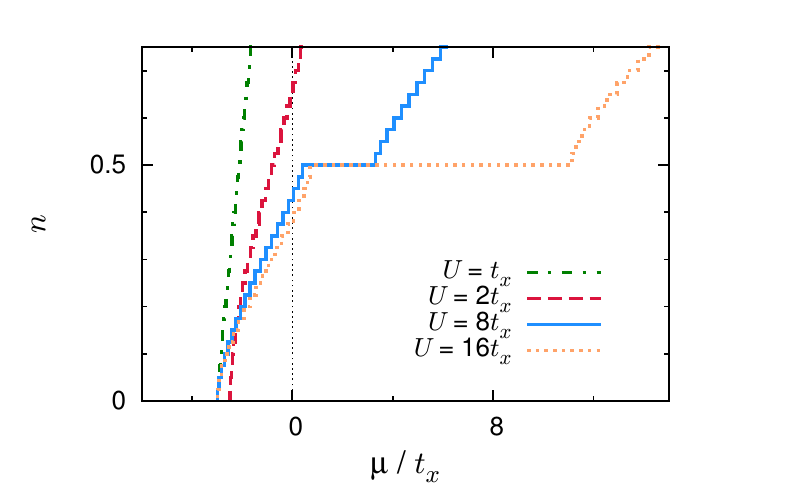}
\caption{
Equation of state $\rho=\rho(\mu)$ for the $SU(2)$ softcore boson model on a $M=2$ leg ladder for several values of $U$.  
}
\label{fig:S5}
\end{figure}

Indeed, in Fig.~\ref{fig:S2} we see that $\Delta_{\rm H}'$ deviates noticeably from $\Delta_H$ for an intermediate range of interactions $U_{01}$. However, both curves are considerably close, such that $\Delta_{\rm H}'$ can be used as a good qualitative measure for the Hall imbalance.

The Hall resistivity $R_{\rm H}$ may be obtained in the spirit of Ref.~\cite{Prelovsek1999} by means of an additional potential term in the Hamiltonian $\mu_y P_y$, which should be adjusted such that $\la P_y \ra = 0$. Instead of calculating higher order energy derivatives as suggested by  Ref.~\cite{Prelovsek1999}, it turns out to be favourable to employ a iterative optimization scheme to determine $\mu_y$. Typically 2 or 3 iteration steps are sufficient to obtain $\mu_y$ with sufficient precision.

%%%%%%%%%%%%%%%%%%%%%%%%%%%%%%%%%%%%%%%%%%%%%%%%%%%%%%%%%%%%%%%%%%%%%%%%%%%%%%%%%%%%%%%%%%
\section{Bose-Hubbard model on a two leg ladder}

In the following we show some details on the softcore Bose-Hubbard model on a $M=2$ leg ladder, corresponding e.g. to the data shown in Figs.~3 of the main text. 
For the cause of numerical simulation we restrict the local Hilbert-space of the model to a low number of bosons per site $n_{max}=4$ which is justified due to the finite repulsive interactions. For strong interactions such Fig.~3 of the main text we keep $n_{max}=3$ and $2$ particles per site.

In Fig.~\ref{fig:S3} and Fig.~\ref{fig:S4} we present detailed data corresponding to Fig.~3 of the main text. Fig.~\ref{fig:S3} shows the finite size PBC results and examples of the extrapolation procedure. In Fig.~\ref{fig:S4} the tDMRG simulation data of the quenched tilt is shown.

In Fig.~\ref{fig:S4} the equation of state for the $SU(2)$ symmetric models is shown which reveals the rich structure of the ground-state even though only a single Zeeman-component is occupied. For strong interactions a series of Mott-insulator plateaus emerges at unit filling $n=1$ and - due to the 2-legs of the model - at half filling $n=0.5$.

\begin{figure}[tb]
\centering
\includegraphics[width=0.48\linewidth]{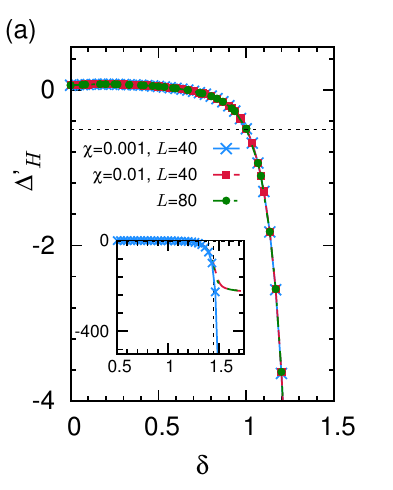}
\includegraphics[width=0.48\linewidth]{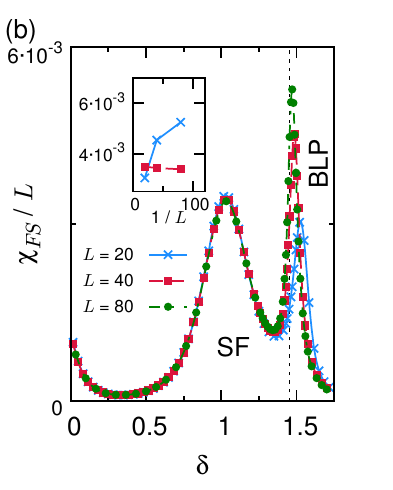}
\caption{
(a) Hall imbalance for the Bose Hubbard model as function of the interaction ratio $\delta=U_{01}/U_{00}$ for $M=2$, $U_{00}=16$, $n=2/3$ (DMRG data). After the phase transition to the BLP phase the Hall imbalance diverges as $\sim 1/\chi$ (see inset).  
(b) Fidelity susceptibility for the same parameters for different . The diverging peak around $\delta\sim 1.3$ indicate the quantum phase transition to the BLP phase. The inset of shows the scaling of the peak height for the QPT (blue curve) and the $SU(2)$ point (red curve).
}
\label{fig:S6}
\end{figure}

Finally, we discuss the phase transition to the biased ladder phase~(BLP), which one may observe for some value $\delta=U_{01}/U_{00} > 1$. In Fig.~\ref{fig:S5} we show $\Delta'_{\rm H}$ for filling $n=2/3$ as function of $\delta$. For $\delta \gtrsim 1.4$ the ratio $\Delta'_{\rm H}$ diverges as $1/\chi$ as shown in the inset of Fig.~\ref{fig:S5}. 
We may further characterize the position of the phase transition by an diverging peak in the fidelity susceptibility~\cite{gu2010} (see Fig.\ref{fig:S6})
\begin{align}
\chi_{FS}(\delta) = \lim_{\delta-\delta' \to 0} \frac{-2 \ln |\langle \Psi_0(\delta) | \Psi_0(\delta') \rangle| }{(\delta-\delta')^2} \,,
\end{align}
with the ground-state wave function $|\Psi_0\rangle$. Interestingly, also at the $SU(2)$-point the model exhibits a local but non-diverging maximum in the $\chi_{FS}$-curve indicating a crossover between different regimes.

\bibliography{references}